\documentclass[12pt]{article}
\pdfoutput=1 
\usepackage{amssymb}
\usepackage{amsmath}
\usepackage{graphicx}

\def\l{\left(}
\def\r{\right)}
\def\la{\langle }
\def\ra{\rangle }

\newcommand{\be}{\begin{equation}}
\newcommand{\ee}{\end{equation}}
\newcommand{\ba}{\begin{align}}
\newcommand{\ea}{\end{align}}
\newcommand{\bg}{\begin{gather}}
\newcommand{\eg}{\end{gather}}
\newcommand{\bseq}{\begin{subequations}}
\newcommand{\eseq}{\end{subequations}}

\title{Fatal youth of the Universe:\\
  black hole threat \\ for the electroweak vacuum during preheating}

\author{Dmitry Gorbunov, Dmitry Levkov, Alexander Panin\\
       {\small\em Institute for Nuclear Research of Russian Academy of
  Sciences, 117312 Moscow,
  Russia}\\  
            {\small\em Moscow Institute of Physics and Technology, 
141700 Dolgoprudny, Russia}
}
\date{}

\begin{document}

\maketitle

\begin{abstract}
  Small evaporating black holes were proposed to be dangerous
  inducing fast decay of the electroweak false vacuum. We observe that
  the flat-spectrum matter perturbations growing at the
  post-inflationary matter dominated stage can produce such black
  holes in a tiny amount  which may nevertheless be sufficient to
  destroy the vacuum in the visible part of the Universe via the
  induced process. If the decay probability in the
  vicinity of Planck-mass black holes was of order one as suggested
  in literature, the absence of such objects in the early Universe
  would put severe constraints  on inflation and subsequent stages
  thus excluding many well-motivated models (e.g.\ the
  $R^2$-inflation) and  supporting the need of new physics in the
  Higgs sector. We give a qualitative 
  argument, however, that exponential suppression of the 
  probability should persist in the limit of small black hole masses. This
  suppression relaxes our cosmological constraints, and, if
  sufficiently strong, may cancel them.
  \end{abstract}
\flushbottom

\section{Introduction and summary}

The electroweak (EW) vacuum of the Standard Model of particle physics
(SM) with top-quark mass, strong coupling constant and Higgs boson
mass taken at the central measured values\,\cite{Olive:2016xmw} is definitely
unstable given the high-order quantum
corrections\,\cite{Bednyakov:2012en,Chetyrkin:2013wya} to the Higgs 
potential. The decay of this vacuum proceeds via tunneling through the
potential barrier to the true vacuum at subplanckian values of the
Higgs field\,\cite{Buttazzo:2013uya,Bednyakov:2015sca}. As a result, a
bubble filled with the Higgs field in the catchment area of the true
vacuum is produced: the field rolls down towards the true vacuum while
the bubble expands occupying more and more space. Fortunately,  this 
catastrophe is found to be extremely rare\,\cite{Buttazzo:2013uya}, so 
that our Universe lifetime grossly exceeds its present age of 14
billion years\,\cite{Olive:2016xmw}.

The late Universe, either dominated by matter or cosmological
constant, is safe for billions of billions of successive human
generations. The early Universe expansion most probably was driven by
some new physics, but the process was arranged in such a way that the
Higgs field had avoided escaping to the true vacuum. This requirement
implies various constraints on the pre-Big-Bang history of the
Universe, including inflation, preheating and reheating stages, which
have been largely discussed in literature, see
e.g.\,\cite{Herranen:2015ima,Ema:2016kpf}. 

Recently it has been suggested\,\cite{Burda:2015isa, Burda:2015yfa,
  Burda:2016mou} that the situation changes completely in the presence
of small evaporating black holes. These objects were proposed to act as
nucleation sites for the bubbles of true vacuum dramatically
increasing the rate of their formation. The largest enhancement was
expected in the case of the smallest-mass black holes which were
argued to kick the Higgs field over the energy barrier and into the
abyss with the probability of order one. Then every black hole at the
last stages of its evaporation  should produce an expanding bubble of
true vacuum around itself. {\em Since we still live in the false
  vacuum, no black hole  had ever completely evaporated during the
  entire history of our Universe.}   

In this paper we turn this observation into model-independent
cosmological bounds and further discuss their possible model-dependent
refinements. Trying to be maximally accurate, we critically 
analyze the black hole induced processes described above. We find that
the proposed effect is most
efficient if the size of the black hole is much smaller than the radius
of the true vacuum bubble forming around it. This apparent violation of
locality rises doubts in physical interpretation of the solutions of
Ref.\;\cite{Burda:2016mou} and suggests that the probability ${\cal 
  P}_{EW}$ of the induced decay, though naturally 
enhanced as compared to the case without black hole, may have been
overestimated in\;\cite{Burda:2016mou}. We therefore keep arbitrary
${\cal P}_{EW}$ throughout the paper.

The
SM provides no mechanisms to produce small black holes
neither today nor in the early Universe, definitely not during the
primordial nucleosynthesis and at the succeeding epochs. Any new
physics operating at earlier times, if it leaves the EW vacuum
metastable, should be of the same kind producing only sufficiently
large black holes which do not evaporate until now. We emphasize that
in this case even cosmological models with primordial black holes
fully evaporated before the beginning of the primordial nucleosynthesis
are excluded.

The earliest time when a black hole might be produced is preheating,
the epoch right after inflation. The cosmological models with copious
black hole production at this stage are rather exotic and have
fine-tuned parameters. In generic models the black holes can be produced,
 though very inefficiently\,\cite{Polnarev:1986bi}.  An example of the 
production mechanism is Jeans instability leading to collapse of
primordial matter perturbations at the post-inflationary matter
dominated stage\,\cite{Khlopov:1980mg}.  In this case 
 strong suppression\,\cite{Polnarev-1,Polnarev-2} of black hole
formation is due to the fact that the typical 
spatial inhomogeneities originating from the flat-spectrum primordial
Gaussian perturbations are too aspherical\,\cite{Doroshkevich} and uneven to
form black holes. 

The main observation of this letter is that {\it in a generic
  inflationary model} with post-inflationary stage of matter
domination continuing long enough for some matter perturbations to
grow and become non-linear, $\delta\rho/\rho\sim 1$, the black holes are
formed in the region occupied by the visible part of the Universe. In fact,  they
  are produced even if the fluctuations grow to
$\delta\rho/\rho\sim 0.1$; in this case a few occasionally largest 
fluctuations of a given wavelength become nonlinear and collapse. Then
transition to the true (subplanckian) vacuum is performed via the black
hole induced mechanism described above. The only general way to
stop black hole formation is early reheating which
reduces the period of post-inflationary matter domination. For
illustration we put a corresponding limit on the reheating temperature
in the large-field inflationary models. Similar limits exist in other
models where the perturbations become nonlinear at preheating. 

In a nutshell, this paper demonstrates that the black holes
  inducing tunneling are important for cosmology. The rate of the
  induced processes, however, should be revised for 
  obtaining robust cosmological constraints.

\section{Long post-inflationary matter domination}
\label{Sec:simple}

Black holes of masses in a wide range can be formed in the early
Universe from collapsing  initial matter density
perturbations. To this end the overdense regions should be squeezed by
gravity within their own Schwarzschild 
radii\,\cite{Zeldovich-1,Hawking:1971ei}.  This happens most
efficiently during matter dominated stage when the pressure preventing
contraction vanishes and the black hole formation is mainly
determined\,\cite{Polnarev:1986bi} by whether the overdense region is
sufficiently spherically symmetric and smooth, or not.

An early matter dominated stage is generically realized in the
inflationary models with massive inflaton. It begins right after
inflation and lasts while the Universe is dominated by the oscillating
inflaton field until reheating. During this stage the Hubble parameter
$H(t)$ and matter (inflaton) density $\rho(t)$ are related by the
Friedmann equation and depend on the scale factor $a(t)$ as
\begin{equation}
  \label{Friedmann}
  H^2(t) = \frac{8\,\pi}{3}\,G\,\rho\propto\frac{1}{a^3(t)}\;,
\end{equation}
where $G$ is the gravitational constant. 
The matter density contrasts with conformal momentum $k$
and subhorizon size  $a/k\simeq R(t)\ll 1/H(t)$ grow linearly with the
scale factor, $\l\delta\rho/\rho\r_k\propto a$, starting from the
primordial value 
$$
\l\delta\rho/\rho\r_{k,\,i}\equiv\delta_{i} \sim 10^{-4}
$$ at the
horizon crossing $R_*=1/H_*$.

Let us begin with the situation when matter domination is long
enough for some of the shortest modes to grow, decouple
from the Hubble flow at turnaround entering the nonlinear regime,
$\l\delta\rho/\rho\r_k\sim 1$,  and then collapse forming clumps of the
inflaton field.  Some of the clumps happen to be sufficiently
spherical and sufficiently smooth to further collapse into black
holes.\footnote{The time scale of this process is about the free-fall
  time in the Tolman solution and hence is of the order of the 
  cosmological time scale determined by the Hubble parameter.} This
process was investigated in
Refs.\,\cite{Khlopov:1980mg,Polnarev-1,Polnarev-2,Polnarev:1986bi},
where the probability of a given clump to be appropriate for
collapsing into a black hole was estimated as
\begin{equation}
  \label{BHprob}
 {\cal P}_{BH}\approx 2\times 10^{-2}\, \l \frac{r_g}{R} \r^{13/2}\;,
\end{equation}
 with $R$ and $r_g$ denoting the size of the clump at
turnaround and the gravitational radius of the resulting black
hole. These quantities are related to the  clump mass $M$ and matter
(inflaton) density 
$\rho$ at turnaround by
\begin{equation}
  \label{clump-mass}
r_g=2 G M\,,\,\;\;\;\;M\approx\frac{4\,\pi}{3}\rho\,R^3\;.
\end{equation}
Taking the matter density from the Friedmann equation
\eqref{Friedmann}, one obtains,
\begin{equation}
  \label{x-ratio}
\frac{r_g}{R}\approx H^2 R^2= \frac{a_*}{a}\simeq\delta_{i}\,,
\end{equation} 
where the second relation accounts for the scale factor dependence of
$H$ and $R$ between the Hubble crossing and turnaround, see
\eqref{Friedmann}. The  third relation uses the fact that the
contrasts grow linearly with the scale factor. Hence, the
probability \eqref{BHprob} is
\begin{equation}
  \label{BHprob-simple-0}
 {\cal P}_{BH}\approx 2\times 10^{-2}\, \delta_{i}^{13/2}\;.
\end{equation}
There are $(HR)^{-3}\simeq \delta_{i}^{-3/2}$
clumps of size $R$ inside the Hubble volume at turnaround, hence, the
probability to have a black hole in that region,
\begin{equation}
  \label{BHprob-simple}
        {\cal P}_{BH, \,hor}\approx 2\times 10^{-22}\,
        \l\frac{\delta_{i}}{10^{-4}}\r^{5}\;,
\end{equation}
is very small.

On the other hand, our present-day Universe with the Hubble parameter
$H_0 \ll H$ has many such regions. Indeed, since their volumes
grow as $a^3$ starting from $H^{-3}$, there are
\begin{equation}
  \label{eq:3}
  N_{hor}= \l \frac{H}{H_0}\r^3 \l\frac{a}{a_0}\r^3
\end{equation}
of them in the visible part of the Universe.

Consider the largest possible black holes formed
right before the reheating: $H = H_{reh}$ and $a = a_{reh}$ in
Eq.~(\ref{eq:3}). At the hot stages the 
entropy in the comoving volume is conserved; parameter $H_{reh}$ is related
to the reheating temperature $T_{reh}$ and the number of
relativistic degrees of freedom $g_{*,reh}$ by the Friedmann equation
$H_{reh}^2\sim G
g_{*,reh} T_{reh}^4$. Then, neglecting some numerical factors, one finds,
\begin{equation}
  \label{eq:4}
  N_{hor}\simeq g_{*,0} \,G^{3/2} \frac{T_0^3}{H_0^3}\,\sqrt{g_{*,reh}}\,T_{reh}^3\,.
\end{equation}
This large number must be multiplied by \eqref{BHprob-simple} to 
estimate the number ${\cal N}_{BH,\, 0}$ of the primordial black holes
{\it within the presently visible part
of the Universe}. Assuming $g_*
\sim 100$, one obtains,
\begin{equation}
  \label{BH-today-present}
        {\cal N}_{BH,\, 0}= N_{hor}\times  {\cal P}_{BH,\, hor}
        \simeq \l\frac{T_{reh}}{3\times 10^{-4}\,\text{GeV}}\r^3
\times \l\frac{\delta_{i}}{10^{-4}}\r^{5}.
\end{equation}
Recall that $a_{reh}/a_{inf} \gtrsim \delta_i^{-1}$ is assumed in this
formula. The number (\ref{BH-today-present})
exceeds unity as far as the reheating temperature is
above MeV scale, which is required for the successful primordial
nucleosynthesis\,\cite{Olive:2016xmw}.  The number of the lighter
  black holes is even larger because the respective perturbations  
start to grow earlier. In particular, the smallest black holes are
formed by the pertubations starting to grow immediately after
inflation. Their number is given by Eq.~(\ref{BH-today-present})
multiplied by $(\delta_i a_{reh} / a_{inf})^{3/2} > 1$.

In cosmological models with realistic preheating temperature
$T_{reh}\gtrsim 100$\,GeV many evaporating black holes are
formed. Recently it was suggested\;\cite{Burda:2015isa,
    Burda:2015yfa, Burda:2016mou} that the electroweak vacuum
decays with enhanced rate\footnote{We consider realistic case of
  small $E_{b}$ as compared to the black hole mass.}
$\mathrm{e}^{-E_b/T_{BH}}$  in the vicinity of small black holes, where
$E_{b} \sim 10^{12}\, \mbox{GeV}$ is the height of the energy barrier
between the vacua and $T_{BH}$ is the Hawking  temperature. This
result is based on Euclidean calculations\,\cite{Burda:2016mou} in the 
Standard Model of particle physics beyond the thin-wall
approximation. It is also supported by the semiclassical exercises in toy 
models describing thin-wall bubbles\,\cite{Parikh:1999mf, Berezin:1999nn,
  Bezrukov:2015ufa, Burda:2015isa, Burda:2015yfa}. From the physical
viewpoint, the result implies that the black hole, like a thermal bath,
activates over-barrier transitions  between the vacua with the Boltzmann
suppression. The smallest black holes at the last stage of their
evaporation have $T_{BH}>E_{b}$ and therefore destroy our vacuum with
the probability of order one. Conversely,  since we still 
live in the metastable vacuum, no black hole had ever completely evaporated
within the visible part of the present Universe.    

As is noted in the Introduction, there is no doubt in qualitative
  role of black holes catalyzing bubble formation. However, the
  applicability of the Boltzmann formula for 
  the decay probability is questionable, as we  explain in 
Sec.~\ref{sec:do-black-holes}.  We therefore denote  by ${\cal
  P}_{EW}$ the probability of induced false vacuum decay during the
last stages of black hole evaporation, the very probability that was
claimed  in literature\;\cite{Burda:2016mou} to be of order one.

One concludes that the EW vacuum is destroyed if 1) it is metastable;
2) the post-inflationary stage is matter-dominated and lasts long
enough for the shortest-scale perturbations to enter the non-linear
regime (i.e.\ $a(t)$ grows by a factor $\delta_{i}^{-1} \sim 10^4$);
3) inside the present horizon the perturbations form more than
  ${\cal P}_{EW}^{-1}$ 
  primordial black holes which are light enough to evaporate down to the 
  planckian masses by now.

Assuming the two first conditions are fulfilled, let us elaborate on
the last condition implying, in particular, that $M_{BH}\lesssim M_c\equiv
10^{14}$\,g, see e.g.\,\cite{Khlopov:2008qy}. The smallest-mass black
holes are formed by the perturbations entering the horizon immediately
after inflation. Computing the mass within the cosmological horizon
$H_{inf}^{-1}$ at that time, we obtain,
\begin{equation}
  \min(M_{BH}) \simeq (2GH_{inf})^{-1}\;,
\end{equation}
where the Friedmann equation~(\ref{Friedmann}) was used. These
smallest black holes are harmless, $M_{BH} > M_c$, if the scale of inflation
is low enough,
\begin{equation}
  \label{eq:7}
  H_{inf} \lesssim (2G M_c)^{-1} \approx
  \mbox{GeV}\;.
\end{equation}
This gives low energy density
\begin{equation}
  \label{eq:1}
  \rho_{inf} \lesssim \frac{3}{32 \pi G^3 M_c^2} \approx (2\times 10^{9} \,
  \mbox{GeV})^4
\end{equation}
at the end of inflation.  Note that the bound (\ref{eq:1}),  \eqref{eq:7}, if
  applicable, is way stronger than the condition $H_{inf}/2\pi
  \lesssim 10^{11}$\,GeV\,, see e.g.\,\cite{Herranen:2015ima, Ema:2016kpf}, 
  ensuring stability of the EW vacuum during inflation.
  
The models obeying \eqref{eq:1} are viable: the primordial black holes
produced at that stage  are harmless until now (yet the catastrophe
awaits us in the future). Recalling the assumption of long enough
post-inflationary matter dominated stage,  one can recast
\eqref{eq:1} as the limit on the reheating scale. In this case
$H_{reh} \leq H_{inf} \delta_i^{3/2}$, and all models with
$M_{BH}>M_c$ have low reheating temperature,
\begin{equation}
  \label{min-temperature}
  T_{reh} \sim \frac{H_{reh}^{1/2}}{(g_* G)^{1/4}} \lesssim 10^{6}\,
  \mbox{GeV}\times\l \frac{\delta_{i}}{10^{-4}}\r^{3/4}\;.
\end{equation}
This inequality can be used together with Eq.~(\ref{eq:1}) to identify
the cosmologically viable models.

On the other hand, if the energy density at the
end of inflation is high and violates~\eqref{eq:1},  some primordial black holes
evaporate before the present epoch and may destroy the EW
vacuum. Requiring the number of black holes (\ref{BH-today-present})
to be smaller than ${\cal P}_{EW}^{-1}$, we 
  obtain a constraint on the reheating temperature,
\begin{equation}
  \label{eq:9}
  T_{reh} \lesssim 3\times 10^{-4} \, \mbox{GeV} \times {\cal
    P}_{EW}^{-1/3} \times \left(
  \frac{\delta_i}{10^{-4}}\right)^{-5/3} \;. 
\end{equation}
If the probability of the induced decay is of order
one, ${\cal P}_{EW} \sim 1$~\cite{Burda:2016mou}, this inequality
excludes all models violating (\ref{eq:1}), (\ref{min-temperature})
and having long enough post-inflationary matter dominated epoch
because the temperature~(\ref{eq:9}) is too low for successful
primordial nucleosynthesis. If  ${\cal P}_{EW} < 1$ but not too small,
such models are still severely constrained by Eq.~(\ref{eq:9}).
The constrained
models include the $R^2$-inflation\,\cite{starobinsky} where the
energy density at inflation is high, the reheating temperature is
low \cite{Vilenkin:1985md,Gorbunov:2010bn}, and the scale factor grows
by $10^7$ between these epochs. Generally, for
the inflationary models with high energy scale, e.g.\ the large-field
models, the only way to avoid the danger is to prevent formation
of the inflaton clumps, so that the scale factor grows by a factor
smaller than $10^4$ during preheating. This places the lower limit on
the reheating temperature,
\begin{equation}
\label{Treh-simple}
T_{reh}\gtrsim 5\times 10^{12} \, \text{GeV}\times
\frac{\rho_{inf}^{1/4}}{ 10^{16}\,\text{GeV}} \times
\left(\frac{\delta_i}{10^{-4}} \right)^{3/4}\,.
\end{equation}
     Viable models satisfy either this inequality or 
       Eq.~(\ref{eq:9}), or
      Eqs.~(\ref{eq:1}), (\ref{min-temperature}). 
    In particular, Eq.~(\ref{Treh-simple}) can be met in inflationary models
with quartic scalar potential and large non-minimal coupling to gravity
similar to the 
Higgs inflation\,\cite{Bezrukov:2007ep} where the reheating
temperature is estimated to be higher\,\cite{Bezrukov:2008ut}.

\section{Beyond the simplest scenario}
In special cases  e.g. when the parameters of a given model~---
reheating temperature, energy density at inflation, duration of the
matter-dominated stage~--- are close to the critical values separating
the EW-safe and EW-destroyed models, one would like to refine our
estimates. In fact, the only situation where the refinement is
definitely needed is
when the post-inflationary matter-dominated stage is long but not
long enough for the shortest perturbations to form clumps with
$\delta\rho/\rho\sim1$ everywhere in the early Universe.  In this 
  situation
the gravitationally bound clumps  still may be formed in a few places due to
positive fluctuations even  if the fluctuations are not yet 
non-linear on average, similar  process happened in the late
Universe when the first stars get ignited.

There are several issues becoming important in this case  of early
reheating. First, one 
naturally concentrates on the perturbations of the shortest
wavelengths which enter the horizon and start to grow immediately
after inflation. However, at the end of inflation both the expansion
law and the scalar perturbation amplitude may deviate (and noticeably,
see e.g.\ the case of the inflaton quartic
potential\,\cite{Jedamzik:2010dq}) from what one has for the reference
modes exiting the horizon some 50-60 e-foldings before the inflation
terminates. Likewise, the expansion right after inflation is not
exactly like at the matter-dominated stage, so that the contrasts do not
immediately approach the linear-with-scale-factor growth.  Hence, it
well may happen in a particular model that it is not the shortest mode
which has the highest amplitude and happens to be the first to approach
$\delta\rho/\rho\sim1$ and enter the non-linear regime.  Finally, the
reheating is also not an instant process, and the expansion  of
  the Universe during this period also departs from that at the pure 
matter-dominated stage slowing down the contrasts growth. All the
aforementioned effects are model-dependent.

Second, to study the features of the inflaton clumps, one has to
express the clump size and height in terms of the relevant parameters
from the inflaton sector. This includes extracting the subhorizon
modes and summing over all shorter-wavelength modes that contribute to the
local spatial inhomogeneity of size $R$. The latter summing
inherits some arbitrariness due to the choice of the window function
which constrains the modes in the space to the region of spatial size
$R$. The common choice is the top-hat filter function which
gives the following dispersion of the density contrasts,  
\begin{equation}
\label{dispersion}
  \la \delta^2_R(t)\ra \equiv \sigma_R^2(t) = \int_{Ha}^{k_{max}}
  \frac{dk}{k}\,{\cal P} (k,t)\times \frac{9\, j_1^2\l R\,k/a\r}{\l
    R\,k/a\r^2}\,, 
\end{equation}
with $j_1(x)$ being the spherical Bessel function and ${\cal P} (k,t)$
representing the matter power spectrum. The integration is performed
over the subhorizon modes, $H\lesssim k/a$, with the upper limit referring to the
shortest modes which exit the horizon at the very end of inflation,
$k_{max}= H_{inf}a_{inf}$. Since the function $j_1(x)$ oscillates
  with decreasing amplitude at large $x$, 
Eq.\;\eqref{dispersion} implies that the modes most relevant
  for the clump formation belong to the interval
${H\lesssim k/a\lesssim 1/R}$.

Third, to estimate the probability of forming a clump of size $R$ out
of perturbations with dispersion $\sigma_R^2(t)$ one can exploit the
Press--Schechter formalism\,\cite{Press:1973iz}. At $\sigma_R^2\ll1$,
 i.e.\ before the perturbations become nonlinear, one finds,
\begin{equation}
  \label{Press-Schechter}
  {\cal P}_{clump} = 
  \int_{\delta_c}^{\infty} \frac{d\delta}{\sqrt{2\pi} \sigma_R} 
  \exp \l -\frac{\delta^2}{2\sigma_R^2} \r \approx
  \frac{\sigma_R}{\sqrt{2\pi}\,\delta_c }
  \exp \l -\frac{\delta_c^2}{2\sigma_R^2} \r
  \;,
\end{equation}
where the threshold value $\delta_c\approx 1.686$ is obtained using
the Tolman solution, see e.g.\,\cite{Gorbunov:2011zzc}. Thus, naively
one would expect to multiply the probability \eqref{BHprob-simple-0}
by the factor \eqref{Press-Schechter} accounting for the fact that
while the perturbations of size $R$ are too small on average, there
are few high fluctuations which might form the clumps.
However, this is not the end
of the story yet.

Forth, the clumps, formed by the modes which are the very first to
become nonlinear, are not entirely homogeneous enough to allow 
black hole formation. Indeed, there are no (or very few)
fluctuations of shorter wavelengths, and the density profile of the
mode itself is not smooth enough to fuel  black hole formation in
the clump center. Therefore, the
probability of finding the smooth enough configuration may receive
additional suppression in this case as compared to Eq.~(\ref{BHprob}).

Let us estimate how early the reheating  must have been
  occurred in order that  
the present visible Universe stays in the electroweak vacuum with high
probability. This means that within the present horizon there are
 {\em less than ${\cal P}_{EW}^{-1}$ regions} which had collapsed into
 black holes during the early post-inflationary stage.
 To make the estimates as general as possible, we avoid including
  some model-dependent effects mentioned above. Namely, we assume
  matter-dominated expansion law keeping in mind that departures from
  this law can be accounted for on model-to-model basis. We also
  keep the dispersion $\sigma_{R}^2$ as a free parameter
  which should be computed via Eq.~(\ref{dispersion}) in a given
  model. The other effects give small impact because they do not
  change the leading exponent in Eq.~(\ref{Press-Schechter}).

The probability of a
perturbation to form a black hole is given by the product of
Eqs.~\eqref{Press-Schechter} and \eqref{BHprob}. Strictly speaking,
this probability should be summed over all spatial scales $R$ permitted
in Eq.~(\ref{dispersion}). However, it is natural to assume that the
leading contribution comes from the smallest perturbations with $R(t) 
\approx a(t)/a_{inf}H_{inf}$ entering the horizon right after
inflation and immediately starting to grow. We consider the case
  when the average height $\sigma_R \simeq a_{reh}
\sigma_{R,\,inf}/a_{inf}$  of these perturbations remains small at
reheating, where $\sigma_{R,\,inf}\sim \delta_i$ is its value
at the end of inflation. Nevertheless,
the highest perturbations collapse
into structures of mass $M \approx (2GH_{inf})^{-1}$. 
Consequently, for the probability to form a
black hole we obtain (instead of \eqref{BHprob-simple-0}),
\begin{equation}
  \label{BH}
 {\cal P}_{BH} \simeq 5\times 10^{-3}\times\sigma_{R,\,inf}
  \l \frac{H_{reh}}{H_{inf}} \r^{11/3}
  \exp \l -\frac{\delta_c^2}{2\sigma_{R,\,inf}^2} 
  \l \frac{H_{reh}}{H_{inf}}\r^{4/3} \r\;.
\end{equation}
Note that although $\sigma_{R,\,inf}\sim \delta_i$, the
numerical factor between these quantities must be estimated
accurately, as it enters the exponent. The probability~(\ref{BH})
accounts for somewhat suppressed amplitude of the initial perturbations
at the smallest scales and for the inhomogeneity of the clump
forming a black hole.

Following the lines of Sec.\,\ref{Sec:simple} one finds for the
probability to have a black hole inside the horizon volume at
reheating (cf. Eq.\,\eqref{BHprob-simple}),
\begin{equation}
  \label{Np}
  {\cal P}_{BH,\, hor} \simeq \frac{{\cal P}_{BH}}{(H_{reh}R)^3} \simeq
  10^{-2}\times\sigma_{R,\,inf} \l \frac{H_{reh}}{H_{inf}} \r^{8/3}
  \exp \l -\frac{\delta_c^2}{2\sigma_{R,\,inf}^2} 
  \l \frac{H_{reh}}{H_{inf}}\r^{4/3} \r.
\end{equation} 
Finally, the number of completely evaporated black holes inside
the visible part of the 
present-day Universe reads,
\begin{equation}
  \label{PBH-refined}
 {\cal N}_{BH,\, 0} \simeq 10^{64}
 \l\!\frac{T_{reh}}{5\times 10^{12}\,\text{GeV}}\!\r^{\!\!\!3} \;
 \l\!\frac{\sigma_{R,\,inf}}{10^{-4}}\!\r \;
  \l\!\frac{H_{reh}}{H_{inf}}\!\r^{\!\!\!8/3}\!\! \;\;
  \exp\! \l \!-\frac{\delta_c^2}{2\sigma_{R,\,inf}^2} 
  \l\! \frac{H_{reh}}{H_{inf}}\!\r^{\!\!4/3} \r\!,
\end{equation}
which replaces Eq.\,\eqref{BH-today-present}. Recall that this
  expression is valid only in the case of relatively short preheating with
  $a_{reh}/a_{inf} \lesssim \sigma_{R,\, inf}^{-1}
  \sim 10^{4}$; in the opposite case one should use constraints from
  Sec.~\ref{Sec:simple}.
As we expected, Eq.~(\ref{PBH-refined}) is exponentially sensitive to the
ratio of inflation and reheating scales $H_{inf}$ and $H_{reh}\simeq
(G g_{*,\, reh})^{1/2}T_{reh}^2$ and weakly depends on
$T_{reh}$ in the prefactor. Requiring
  that no more than ${\cal P}_{EW}^{-1}$ black holes are formed,
  ${\cal N}_{BH,\, 0} {\cal P}_{EW} < 1$,
  we express $H_{reh}/H_{inf}$ from this inequality 
and obtain a constraint
\begin{equation}
  \label{eq:6}
  \frac{H_{reh}}{H_{inf}} \gtrsim 3\times 10^{-5} \times
  \left(\frac{\sigma_{R,inf}}{10^{-4}} \right)^{3/2}\;, \qquad \qquad
  \frac{a_{reh}}{a_{inf}} \lesssim 10^3 \times
  \left(\frac{10^{-4}}{\sigma_{R,inf}}\right) \;,
\end{equation}
 where $\sigma_{R,\,   inf} \sim \delta_i \sim 10^{-4}$ and $T_{reh}
\sim 5\times 10^{12} \, \mbox{GeV}$ are used in the prefactor of
Eq.~(\ref{PBH-refined}) in accordance with Eq.~(\ref{Treh-simple})
and we assume that ${\cal P}_{EW}$ is not too small.
Equations~\eqref{eq:6} 
slightly refine the naive condition $a_{reh}/a_{inf} \lesssim
10^4$ of no black hole formation and the respective
constraint (\ref{Treh-simple}) on the reheating temperature which
takes the form 
\begin{equation}
\label{eq:8}
T_{reh}\gtrsim 3\times 10^{13} \, \text{GeV}\times
\frac{\rho_{inf}^{1/4}}{ 10^{16}\,\text{GeV}} \times
\left(\frac{\sigma_{R,inf}}{10^{-4}} \right)^{3/4}\;.
\end{equation}
The models violating this inequality or Eq.~(\ref{eq:6}) are excluded.

Constraints (\ref{eq:6}) are valid if ${\cal P}_{EW}$ is larger
  than the critical value
  \begin{equation}
    \label{eq:11}
          {\cal P}_{EW,\, c} \sim 10^{-48}
           \l\!\frac{T_{reh}}{5\times 10^{12}\,\text{GeV}}\!\r^{-3} \;
           \l\!\frac{\sigma_{R,\,inf}}{10^{-4}}\!\r^{-5}
  \end{equation}
which one can read off (\ref{PBH-refined}). Above this value ${\cal
  P}_{EW}$ enters logarithmically into the constraints.
  One concludes that the black holes are produced if
  the Universe stretches $10^{3}$ times (\ref{eq:6}) during preheating and the
  decay probability is not exceedingly low, ${\cal P}_{EW}> {\cal
    P}_{EW,\, c}$. 
\section{Do small black holes catalyze the EW vacuum decay?}
\label{sec:do-black-holes}

As we see, the danger of black hole induced false vacuum
decay imposes severe constraints on 
the inflationary models. Thus, it is natural to pay detailed attention to
this process.
In Sec.~\ref{Sec:simple} we mentioned the arguments of
Ref.\,\cite{Burda:2016mou} suggesting that the probability 
  to form a
bubble of true vacuum around an isolated black hole of mass $M_{BH}$ is
suppressed by the loss of black hole entropy $\Delta S$ in the
process, ${\cal P}_{EW} \propto  \mathrm{e}^{-\Delta S}$. If the bubble
mass  $E_{b}\equiv \Delta M_{BH}$ is much smaller than $M_{BH}$,
this suppression reduces to the Boltzmann factor 
\begin{equation} 
  \label{eq:2}
  {\cal P}_{EW} \propto \mathrm{e}^{-E_b/T_H}\;, 
\end{equation}
involving the black hole temperature $T_{H} \equiv (8\pi G
M_{BH})^{-1}$. Then the transition to the true vacuum becomes
unsuppressed at 
\begin{equation}
  \label{eq:5}
  T_{H} \geq E_b\;,
\end{equation}
indeed implying that the black holes of sufficiently small mass
catalyze 
decay of the EW vacuum.
  
Note, however, that the result\;\eqref{eq:2} of Ref.\,\cite{Burda:2016mou} 
essentially relies on  the
interpretation of static critical bubbles surrounding the black
holes 
as Euclidean instantons describing EW vacuum decay. To understand
these  solutions physically, consider the regime ${E_b \sim \mbox{few}
\times T_H}$ when the probability (\ref{eq:2}) is relatively large yet
exponentially suppressed, so that the semiclassical methods are
applicable. In  this case the  size and mass  $r_g,
M_{BH}$ of the black hole and the respective parameters $R_b$, $E_b$ of the surrounding
bubble are essentially 
different. Indeed, $M_{BH}/E_b \sim  
G M_{BH}^2 \gg 1$. Thus, back-reaction of the bubble on the
background geometry  is small. On the other hand, $R_b E_b \gg 1$
because the bubble is classical, and therefore $r_g/R_b \ll E_b/T_H
\sim 1$. This means that the major part of the true vacuum bubble
lives in flat spacetime far away from the black hole. As a result, the
solution of Ref.\,\cite{Burda:2016mou} coincides with the flat-space critical bubble up to
small corrections.

Given the flat-space nature of the solutions in \cite{Burda:2016mou}
at $E_b \sim T_H$, the drastic enhancement of the related
probabilities looks surprising. Technically, the difference is related
to the fact that the Wick-rotated Schwarzschild time $\tau = it$ is
periodic i.e.\ takes values on the circle ${0<\tau \leq T_H^{-1}}$. This
property holds even in the spatial regions far away from the black hole
where the spacetime is flat. As a consequence, the static bubbles of
energy $E_b$ and any size have finite action $B = E_b/T_{H}$ and give
contributions (\ref{eq:2}) to the Euclidean path integral. Likewise,
the correlator of the quantum field $h(x)$ in the Euclidean
Schwarzschild background coincides with the thermal correlator rather
than with the vacuum one, even if it is computed far away from the
black hole. Since we cannot pretend that the black hole changes
physics in the distant parts of the Universe, the Euclidean
Schwarzschild spacetime should be interpreted as describing a black
hole surrounded by the infinite bath of temperature $T_{H}$
\cite{Hartle:1976tp}
rather than an isolated black hole in empty spacetime. Then
the solutions obtained in \cite{Burda:2016mou} give the rate\footnote{Note
  that the quantum corrections to the Higgs potential should be computed
  on the same background with periodic $\tau$ as the leading-order
  semiclassical solutions. This procedure produces finite-temperature
  Higgs potential rather than the vacuum one. The probability of the
  EW vacuum decay receives additional suppression in this case
  \cite{Salvio:2016mvj}.}~(\ref{eq:2}) of false vacuum decay activated
by fluctuations in the infinite-size thermal bath of temperature
$T_H$. In the regime (\ref{eq:5}) this process is sensitive to
  the bath itself rather than to the tiny black hole in the bubble center.

In reality, small-mass black holes are not isolated but surrounded
by their own Hawking flux of temperature $T_H$. The energy density
within this flux, however, decreases as $r^{-2}$ and becomes
essentially lower than thermal at the distances of several
Schwarzschild radii. The probability~(\ref{eq:2}) is not applicable to
false vacuum decay near such black holes unless their sizes are
comparable to the sizes of the true vacuum bubbles, $r_g \sim
R_b$. While in the latter case one expects to find
enhancement\footnote{In particular, due to smaller bubble energy in
  the black hole gravitational well~\cite{Tetradis:2016vqb}.}, no
unsuppressed vacuum decay should occur in the regime 
(\ref{eq:5}). 

A question remains, however. Common knowledge suggests that 
black holes spit various field configurations, in particular, the
bubbles of true vacuum, with Boltzmann-suppressed probability
(\ref{eq:2}). This intuition is based on numerous semiclassical
exercises with thin-wall bubbles, see
e.g.\,Refs.\,\cite{Parikh:1999mf, Berezin:1999nn,
  Bezrukov:2015ufa}, 
which did not rely on the Euclidean methods at all. One therefore can
imagine a dynamical process where the high-temperature black hole
produces a Higgs field bubble of initial size $r_g$. The bubble  expands with
nonzero velocity and eventually reaches the critical size $R_b \gg
r_g$. However, the Higgs field bubble in the SM has thick walls,
it is formed by the 
field quanta with typical wavelength $R_b$. A configuration of this
kind cannot originate from the local area of small size $r_g$ in the
course of classical evolution. One therefore expects the probability
of producing such a bubble to be exponentially
suppressed as compared to Eq.~(\ref{eq:2}).

To get a quantitative feeling of the suppression, we
  recall that the probability of emitting a particle of wavelength
  $R_b \gg r_g$ from the black hole involves, in addition to the thermal rate,
  a gray factor $\Gamma \propto (r_g/R_b)^2$ \cite{ST:CH,Page:1976df}.
  The bubble of true vacuum containing $E_bR_b \sim 10^3$ quanta is
  therefore produced with the probability suppression $\Gamma^{E_b
    R_b} \propto(r_g/R_b)^{{10^3}}$ in addition to Eq.~(\ref{eq:2}).

Depending on its size, this  additional suppression does or does not 
make meaningless the cosmological bounds of the type
considered in this paper. In particular,
Eq.~(\ref{BH-today-present}) shows that even in models with relatively
low reheating temperature $T_{reh} \sim 10^9$\,GeV like the $R^2$
inflation\,\cite{Vilenkin:1985md, Gorbunov:2010bn} the number of
primordial black holes in the visible part of the
Universe~(\ref{BH-today-present})  is relatively large, ${\cal
  N}_{BH,\, 0} \gtrsim  10^{37}$. When multiplied by the exponentially
suppressed probability ${\cal P}_{EW}$ of producing a bubble near
each black hole,  this number  is still larger than one if ${\cal
  P}_{EW}>{\cal P}_{EW,\, c}$, cf.\ Eq.~(\ref{eq:11}). Then the
respective model (i.e.\  the $R^2$-inflation in our example) is excluded. 

\section{Conclusion}
\begin{figure}[t]
  \centerline{\includegraphics[width=.7\textwidth]{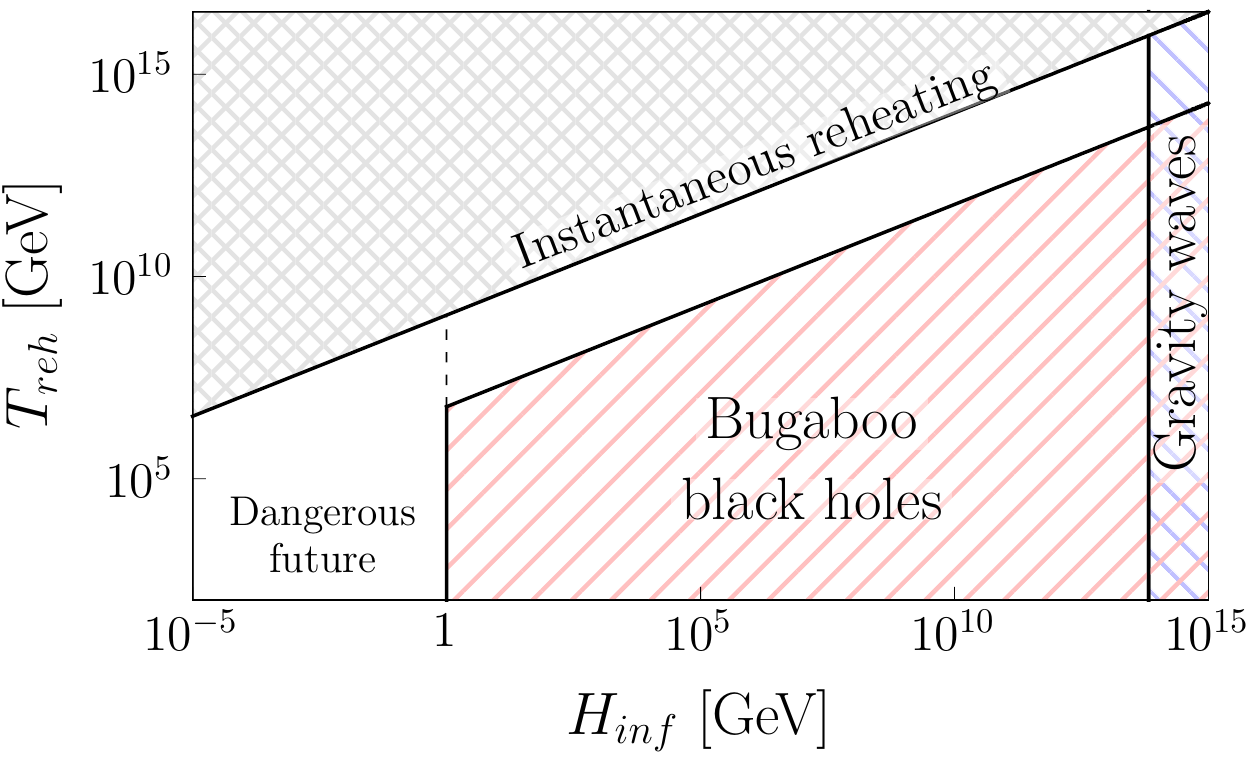}}
  \caption{Limits on the cosmological models with inflation scale
    $H_{inf}$ and reheating  temperature $T_{reh}$. If the probability
    of induced vacuum decay is not too small, ${\cal P}_{EW} > {\cal
      P}_{EW,\, c}$, the value of $T_{reh}$ cannot be
    much lower than  the temperature of the instantaneous reheating,
    see Eqs.~\eqref{eq:6}. This leaves  the narrow allowed (white) strip in the
    $(H_{inf},\; T_{reh})$ plane overlapping with the  experimentally allowed
    region $H_{inf}\lesssim 7\times 10^{13}\,
    \mbox{GeV}$~\cite{Ade:2015tva, Array:2015xqh}. The second allowed
    (white) region in the lower left corner of the plot represents
    models with large primordial black holes which do not evaporate
    until now, see Eq.~(\ref{eq:7}). In that case the danger awaits us in the
    future.\label{fig:limits}}
\end{figure}
To summarize, recently suggested process of black hole induced false
vacuum decay may exclude generic inflationary models with
sufficiently long post-inflationary matter-dominated stages because the
inflaton inhomogeneities grow, decouple from the 
Hubble flow, and a few of them form black holes catalyzing
decay of the EW vacuum. The only general way to suppress
  this black hole formation is early reheating which stops
  gravitational contraction of the matter perturbations by
  nonzero pressure. Then the condition of having short enough
  preheating stage severely constrains the inflationary models and
  reheating mechanisms\footnote{More accurately, this bound
  constrains duration of the matter-dominated epoch until production
  of  relativistic particles which stop perturbation growth and black
  hole formation. By itself, thermalization is not needed for the
    radiation dominated stage to settle.}, see
  Fig.~\ref{fig:limits}. Similar constraints exist 
  in models with other than matter dominated expansion
  laws at preheating if the contrasts of matter perturbations 
  grow sufficiently fast at this stage to approach the non-linear
  regime.  
If the constraints are not met, the scalar sector of the
SM should be modified in a way 
to make the EW vacuum true, and that gives one more argument in favor of
new physics in the Higgs sector. As we also argue, the
process\;\cite{Burda:2016mou} of black hole induced tunneling deserves 
further investigation, since the physics underlying it remains
hidden and its probability may have been overestimated. 

We thank V.~Rubakov, P.~Satunin, and S.~Sibiryakov for discussions.
This work was supported by the grant RSF
16-12-10494. D.L. thanks CERN for hospitality.


\end{document}